\documentclass[12pt]{article}
\usepackage{graphicx}
\usepackage{amssymb,amsmath,amsfonts,palatino,amsthm}
\usepackage{amssymb}
\usepackage{epstopdf}
\usepackage{slashed} 
\newcommand{\f}{\begin{equation}}
\newcommand{\ff}{\end{equation}}

\begin{document}

\title{The dynamics of difference \\}

\author{Lee Smolin\thanks{lsmolin@perimeterinstitute.ca} 
\\
\\
Perimeter Institute for Theoretical Physics,\\
31 Caroline Street North, Waterloo, Ontario N2J 2Y5, Canada}
\date{\today}
\maketitle
\begin{abstract}
A proposal is made for a fundamental theory,  in which the history of the universe is constituted of views of itself.  Views are attributes of events, and the theory's only be-ables; they comprise information about energy and momentum transferred to an event from its causal past.  
A dynamics is proposed for a universe constituted of views of events, which combines the energetic causal set dynamics with a potential energy based on a measure of the distinctiveness of the views, called the variety\cite{variety1}.
As in the real ensemble formulation of quantum mechanics\cite{real2}, quantum pure states are associated to ensembles of similar events; the quantum potential of Bohm then arises from the variety. 
\end{abstract}
\newpage
\tableofcontents


\section{Introduction}

In this paper, I would like to propose a new ontology for physics, in which the history of the universe is constituted of views of itself.   A view is, roughly, what would be seen from an event.  More precisely, in a particle formulation, the view of an event is the set of incoming energy-momentum vectors which coincide or interact at the event.  These can be represented as a punctured two-sphere, with labels.  The position of a puncture on the $S^2$ represents the incident direction of a photon, or other particle, incoming to an event, while the label represents its energy.

An event also has a set of outgoing energy-momentum vectors which convey energy-momentum to other events. These constitute the causal future of the event.  Hence, the 
transmissions of energy-momentum  generate a causal 
structure\footnote{Causal set models of quantum spacetime were introduced in\cite{cs}.
Energetic causal set models differ from other spacetime-free causal set approaches, e.g. \cite{Fotini1} proposed causal sets based on quantum information processing systems, and ref \cite{Cohl} proposed causal sets constructed out of standard model particles.}.   
One name for the theory to be presented here is then a {\it causal theory of views.}

Along with a new ontology, I propose a new framework for a fundamental theory of physics.  In it, gravity is not quantized, rather we propose a common completion for quantum mechanics and general relativity.   Here we show how non-relativistic quantum mechanics emerges in a suitable limit.  In previous papers we have show how a theory of interacting particles, moving in an emergent Minkowski spacetime, arises\cite{ECS1,ECS2,ECS3}.  The remaining step of showing how general relativity emerges is saved for a later paper.  

This proposal is inspired by Leibniz's Monadology, and is also a combination and continuation of three recent developments: the energetic causal set model\cite{ECS1}-\cite{ECS4} we developed with Marina Cortes, relative locality\cite{rl1,rl2}, which we developed with Amelino-Camelia, Freidel and Kowalski-Glickmann, and the real ensemble approach to quantum mechanics\cite{real1,real2}.   In different ways each of these three approaches expresses a common foundation in relationalism and the hypothesis that 
time is fundamental and irreversible\cite{SURT,TR}, but there is a more specific  connection, which will be spelled out shortly.    

The fact that the views are described in terms of punctured two-spheres  makes contact with loop quantum gravity, where Hilbert spaces of observables that live on boundaries of quantum spacetime live on punctured two-spheres\cite{linking}. This connection will  be discussed elsewhere, based on the construction of spin foam models which are causal sets\cite{wolfgang,ECS3}.

The central notion in physics up to this time has been locality, but there are two indications that it may not continue to be a primary notion in the quantum theory of gravity.  

\begin{itemize}

\item{}The experimental disconfirmations of the Bell inequalities tell us that any completion of quantum mechanics that gives us a complete description of individual events will involve explicit non-local interactions.

\item{}Indications from quantum gravity suggest that there is a more fundamental level of description in which there is causal structure but no space.  Instead, space emerges as a low energy description of nature, and with it emerges locality.


\end{itemize}

I  therefor propose to replace and subsume locality in spacetime with a deeper notion, which  is similarity of views.  This idea comes from the real ensemble 
formulation of quantum mechanics\cite{real1,real2}, but the emphasis on a causal structure among the views merges that approach with relative locality and energetic causal sets. 

Leibniz in his Monodology described a relational universe built on a plurality of views of the world, each from the perspective of a different fundamental entity.  The present proposal is a modern instantiation of that vision.

Here are the basic ideas we introduce or develop here:

\begin{itemize}

\item{}The history of the world is made from events.  
Events have predecessors, which are a set of past events that transferred energy and momentum to it, and decendents, to whom they transfer their energy and momenta.  The transferring  of energy and momentum defines a causal structure.
${\cal P}ast (I)$ and 
${\cal F}ut (I)$ are the events in the immediate causal past and causal future of $I$. 

\item{}Each event, $I$,  has a view of the world, from its perspective, ${\cal V}_I$, which is a compendium of information that arrives at the event from its causal past. 

\item{}We endorse Leibniz's principle of the identity of the indiscernible (PII), according to which  each event is completely and uniquely characterized by its view.  By uniquely, we mean that the universe is generated by a dynamics which ensures that no two events in the history of the universe have the same view.

\item{}Thus, the PII Is implemented dynamically via an interaction that drives all pairs of views to differ.  We make use of the notion of the variety of a system\cite{variety1}, 
which we developed a long time ago with Julian Barbour. 
The variety, $\cal V$ of a complex system is a measure of the diversity among the views of that system from different constituents.  We identify the potential energy
of our system of views with the negative of their variety, a hypothesis we introduced in \cite{real1}.

\item{}The views do not live in spacetime.  They live rather in products of momentum spaces, as a view is made up of incoming energy-momentum.  
The fact that we perceive our past
as a set of incoming energy-momenta does not commit ourselves to the expectation that the universe is a lorentzian spacetime.  This approach is, in a way, Kantian, in that the apparent $3+1$ dimensionality and lorentz invariance of our perceived world reflects the structures through which we perceive the world-the views-and are not necessarily realized as properties of the world itself. 

\item{}This formulation is a discrete instantiation of the idea of relative locality\cite{rl1,rl2}, according to which spacetime is reconstructed from information about incoming energy-momentum.  A reconstruction will have defects, and one is that the locality of distant events is dependent both on the observer and the energy carried by the probe that observer uses to image the event.   At a sufficient level of coarse graining there may be, for each observer and probe energy, a separate emergent spacetime into which the events may be embedded, respecting their causal relations.  A certain limit of the energetic causal set dynamics defines a relative locality model.

\item{}In the absence of a fundamental spacetime distance there is a metric on the views,
denoted $h(I,J)$, which measures their similarity. {\it  Two events are nearby in the space of views if they have similar views of their causal pasts.}

\item{}Locality in  $h(I,J)$ replaces spacetime locality.  We specify dynamics according to which two events are likely to interact if their view of their pasts are similar, i.e. if their similarity distance,  $h(I,J)$, is small.  These similarity interactions need not respect the causal structure;  neither the fundamental causal structure or the causal structure of an emergent spacetime.

\item{} This concept of similarity distance subsumes ordinary distance because, if two events are nearby in spacetime, they have similar causal pasts. The hypothesis that interactions are  local in similarity explains the concept of local interactions in spacetime.

\item{}But elementary events may sometimes be similar, even when they are far away in the emergent spacetime\cite{real1,real2}.  This, I propose, is the origin of the apparent non-locality in quantum physics.  That is, events that have similar views, that are nearby in the metric on the space of views, are likely to interact.  This is true whether their views are similar because they are nearby in the causal structure, or not.

\item{}Sufficiently complex events, whose views take much information to describe, are likely to be highly unique, and hence isolated in the space of views. But a simple event, whose view has few attributes, will likely have a view similar to those of many other events scattered through the universe.  Thus, we have a natural reason to expect that notions of simplicity and complexity come into the dynamics of a system of views.

\item{}In particular, many simple events can be grouped into ensembles of sufficiently similar events.  The elements in these ensembles interact with each other, without regard to locality in the emergent spacetime.  These ensembles are what the quantum state describes.  This is the basic idea of the real ensemble formulation of quantum mechanics\cite{real1,real2}.   Below, in section 4, we indicate how the non-relativistic formulation of quantum mechanics based on the real ensemble formulation presented in \cite{real2} arises from the non-relativistic limit of the theory presented here.  In particular, as we showed in \cite{real2} the variety is, in the limit in which the ensembles are large, well approximated by the integral of the Bohmian
quantum potential.

\end{itemize}

More specifically:

\begin{itemize}

\item{}The view of an event, $I$, consists of a set of energy-momentum vectors, $[p_I^K]_a$ which are the quanta whose interaction generates the event.  These live in a momentum space, $\cal P$, about which we will have more to say below.  The $[p_I^K]_a$
is the energy momentum received from event $K \in {\cal P}ast (I)$.    Thus, the world is an example of an energetic causal set\cite{ECS1}-\cite{ECS4}.

\item{}The  $[p_I^K]_a$ are
represented by points on the two-sphere indicating direction, each labeled by a number indicating the incoming energy.  

\item{}Momentum space may be flat, as in special relativity, or curved, as is explored in models of relative locality\cite{rl1,rl2}.  In the case that the momentum space is maximally symmetric, a copy of the local lorentz group acts separately on each ${\cal V}_I$ as $SL(2,C)$; this is a local gauge symmetry. 

\item{}Outgoing energy momenta from an event $I$ to an event $K$ in its immediate causal future are labeled by $[q_K^I]_a \in {\cal P}$.  This is related to the momentum as received by $K$, which is $[p_I^K]_a$ by a lorentz transformation,
\f
[p_I^K ]_a = [U_I^K ]_a^{\ \ b} [q_K^I]_b
\label{pt}
\ff
These transformations underlie the curvature of the emergent spacetime.

\end{itemize}

\section{Kinematics of energetic causal sets}

I then postulate that the universe is an energetic causal set\cite{ECS1}, consisting of events, $E_I$ and causal relations $R_{IJ}$, where $J<I$.  Each $E_I$ has a past set ${\cal P}ast_I$
consisting of events which directly causally proceed it\footnote{For more details of the construction of energetic causal sets, see \cite{ECS1,ECS2,ECS3}.}.   

Associated with each immediate relation $<JI>$  are two future pointing momenta, one transmitted from $J$, the second received by I, which we will call $q_{a IJ}$ and $p_{a IJ}$.  
These are related by (\ref{pt}).
We assume a metric on momentum space, $g^{ab}$ and impose the energy momentum relations.
\f
\tilde{\cal C}_{IJ} =q_{a IJ}q_{b IJ} g^{ab} + m^2 c^4 = 0, \ \ \ \ \ {\cal C}_{IJ}= p_{a IJ}p_{b IJ} g^{ab}
+ m^2 c^4 =0  .
\ff


The view of an event is  a set of labeled points on an $S^2$, corresponding to its past set,
with a point $\sigma_J \in S^2$ labeled by the energy transmitted $e_{Ij}=p_{0 IJ}$.
This follows by a standard transformation between the space of directions and points on an $S^2$.

A given event ,  $E_I$ gives rise to a number $d_I$ of descendent events, ${\cal F}ut_I$,  and so is represented as labeled points in the $d_I$ views of these descendants. Let's call these
$q_{a K}^I$.  At each event we have a conservation law,
\f
{\bf P}_a^I = \sum_{J \in {\cal P}ast_I} p_{a J}^{I} = {\bf Q}_a^I = \sum_{K \in {\cal F}ut_I} q_{a K}^{I} 
\ff
This is generated by a constraint,
\f
{\cal P}_a^I = {\bf P}_a^I - {\bf Q}_a^I =0
\ff
Note that ${\bf P}_a^I$ is timeline and future pointing.  

\section{Dynamics of energetic causal sets}

We combine the dynamics from energetic causal sets with those from the real ensemble formulation.  The dynamics is specific in two steps.

\begin{itemize}

\item{}There is a process which acts to construct the causal set of events, by continually creating the descendants of present events.   One hypothesis for this event creation process was described, and studied in \cite{ECS1}.  This process weighs the likelihood of the creation of an event by the uniqueness of its causal past, or view.

\item{}Energy and momentum are propagated on the growing causal set by equations of motion, which come from varying an action.

This action is a sum of terms,
\f
S= S^{ECS} + S^{RE}
\ff
where the first term $S^{ECS}$ comes from energetic causal set models\cite{ECS1,ECS2}.  
The second term in the action, $S^{RE}$ comes from the real ensemble approach to quantum mechanics\cite{real2} and is defined in terms of a notion of variety in (\ref{SRE}) below.

\end{itemize}

Thus, in detail\cite{ECS1}
\begin{eqnarray}
S^{ECS} & = &  \sum_I  \sum_{J \in {\cal P}ast_I} \left [  N ( p_{a I}^J p_{b I}^J g^{ab} +m^2 c^2 )
+ \tilde{N} (q_{a J}^I q_{b J}^I g^{ab}  +m^2 c^2 ) + w^{a I}_J (p_{aI}^J - [U_I^J]_a^b q_{bJ}^I  )
\right ]
\nonumber \\
&& + \sum_I z^a_I [{\cal P}_a^I  ]
\end{eqnarray}

We see that the Lagrange multipliers $z^a_I$ and $w^{a I}_J$ have dimensions of length, while $N$ and $\tilde{N}$ have dimensions of time over momentum (which we can write as
$\frac{\hbar}{m^2 c^2}$ for some mass $m$).

\section{The dynamics of maximal variety}

In this paper we are mainly concerned with the second stage of the dynamics.  We assume the causal set has  been determined and investigate the consequences of extremizing the action to determine the distribution of energy and momentum transferred in the various causal processes.  The chief novelty is the role of the potential energy related to the variety, $S^{RE}$.

The second term in the action $S^{RE}$, comes from the real ensemble formulation\cite{real2}, where it measures the variety among the views.  This is defined as follows.  We first define the {\it distinctiveness} of two elements, $I$ and $J$ to be a measure of the differences between the views of $I$ and $J$.  
This is denoted ${\cal D }_{IJ}$.

To construct the distinctiveness we have to consider the view of an event in its own reference frame.  Each event, $I$, has a centre of mass frame given by
\f
{\bf P}_a^I = \sum_{K \in {\cal P}ast (I)} p_{a I}^{ \ \ K}
\ff
This reference frame is defined by a unit timelike vector
\f
{\bf n}_a^I = \frac{{\bold P}_a^I }{|-{\bold P}_a^I |}
\ff
and ``spatial" metric
\f
h^I_{ab} = g_{ab} + {\bold n}_a^I {\bold n}_b^I 
\ff
so that, 
\f
{\bold n}_a^I h^{ab}=0,  \ \ \ \ {\bold n}_a^I {\bold n}_b^I g^{ab}=-1
\ff

Each incoming energy-momenta may be converted into an energy $E_I^K$ and spatial
momenta $R_{a I}^{\ \ K}$ by
\f
E_I^K = {\bold n}^a_I p_{a I}^{ \ \ K}, \ \ R_{a I}^{ \ \ K} = h_a^b p_{b I}^{ \ \ K}
\ff

The distinctiveness, ${\cal D }_{IJ}$, is defined when $d_I=d_J$ and is the best matching of the two views, each in their centre of mass frames, modulo rotations and permutations.
\f
{\cal D}_{IJ}=\frac{1}{N} <
\sum_{K}  \left [ \frac{R_{a I}^K}{| R_{a I}^K |}   - \frac{R_{a J}^K}{| R_{a J}^K | }     \right   ]^2 >_{\mbox{best matching over rotations and permutations}}
\ff
I will define best matching below.

The distinctiveness is large when the two views are very different and approaches zero when they become more similar.  Smaller spatial momenta are weighed more heavily in the comparison of views.  Hence, the distinctiveness  defines a metric on momentum space, given by
\f
h_{IJ} = {\cal D}(I,J)
\ff

The variety is then defined to measure the distinguishability of all the elements from each other.
\f
{\cal V} = \frac{1}{N(N-1)} \sum_{I \neq J} {\cal D}(I,J)
\ff
The variety is large when many views are distinct.

We take the potential energy to be proportional to the negative of the variety
\f
S^{RE} =  g {\cal V}
\label{SRE}
\ff
where $g$ is a coupling constant to be determined.  Minimizing the potential energy then
corresponds to maximizing the variety.

\subsection{Best matching}

The notion of best matching comparisons comes from the work of Babour and Bertotti on relational dynamics\cite{best}.  We define a rotation $[{\cal O}_I^J]_i^j \in SO(3)$; 
one for every pair of events, $I$ and $J$.  Then, to make the formulas simple we define,
\f
r_I^{j i} =\frac{R_{a I}^K}{| R_{a I}^K |}
\ff 
The best matched difference over rotations is the minimum of
\f
{\cal D}_{IJ}=\frac{1}{N} <
\sum_{K}  \left [ r_{i I}^K   - [{\cal O}_I^J]_i^j r_{j J}^K \right   ]^2 >
\label{DIJ}
\ff
over variation of the $[{\cal O}_I^J]_i^j$'s.

This introduces a gauge field $[{\cal O}_I^J]_i^j$ over the complete graph whose nodes are
events, which is reminiscent of quantum graphite models\cite{qgraph}.  If we work to higher order, the fluctuations will induce a dynamics for the $[\tilde{\cal O}_I^J]_i^j$.  We can define an
expansion around the best matched values, $[\tilde{\cal O}_I^J]_i^j$
\f
[{\cal O}_I^J]_i^k= [\tilde{\cal O}_I^J]_i^j \left ( \delta_j^k + [{\cal A}_I^J ]_j^k  \right ) + \dots
\ff
In the Newtonian (non-relativistic) limit we can neglect these spacetime gauge fields,
$ [{\cal A}_I^J ]_j^k$.

\subsection{Equations of motion}

The variation by the lagrange multipliers $\cal N$ and $\tilde{\cal N}$ yield the energy momentum relations
\f
{\cal C}^I_K = \frac{1}{2} h^{ab} p_{a K}^I p_{b K}^I  + m^2 c^4=  0 , \ \ \ \ \  \tilde{\cal C}^I_K = \frac{1}{2} h^{ab} q_{a K}^I q_{b K}^I + m^2 c^4 =  0 .
\label{em}
\ff
We next impose
 the equations of motion got by varying the $w^{aI}_J$ lagrange multipliers 
\f
p_{aI}^J - [U_I^J]_a^b q_{bJ}^I=0.
\ff
In this paper, I will assume all he parallel transports $ [U_I^J]_a^b =\delta_a^b$ 
and so the result is just to set
\f
p_{aI}^J =q_{aJ}^I
\ff

The variation by $p_{a I}^K$ then yields
\f
{\cal M}_I^K p_{a I}^K  h^{ab}= z^b_I -z^b_K + g {\cal F}_{I a}^K 
\label{emergence}
\ff
where ${\cal M}_I^K ={\cal N}_I^K  + \tilde{\cal N}_I^K$  and $ {\cal F}_{IK}^a$ is the {\it quantum force},
\f
 {\cal F}_{I a}^{\ K} = \frac{\partial {\cal V}}{\partial p_{a I}^K }
\ff

\subsection{Non-relativistic limit of the variety }

We are interested in the non-relativistic limit.  To find it, we first project the equation of motion into the spatial momenta of the event $I$,
by multiplying by $h_a^{I \ b}$.
\f
R_{a I}^{ \ K} = P_{a I}^K  h^{I a}_b= \frac{1}{{\cal M}_I^K }  h^{I a}_b (z^b_I -z^b_K  )+ g {\cal F}_{I a}^{\ K}  
\ff
We neglect the quantum force, and pick the gauge in which the lapses are unity,
\f
{\cal M}_I^K = 1
\ff
We then expand in powers of $g$.
\f
R_{a I}^{ \ K}= R_{a I}^{ 0\ K} + g R_{a I}^{1 \ K}+ O(g^2)
\ff
\f
p_{a I}^{ \ K}= p_{a I}^{ 0\ K} + g p_{a I}^{1 \ K} + O(g^2)
\ff
and so on.

Then we have the spatial difference of the $I$'th event with the $K$'th event, as seen in the $I$'th reference frame.
\f
R_{a I}^{0 \ K} =  h^{I a}_b (z^b_I -z^b_K  )
\ff

In the non-relativistic limit we may assume that to a good approximation the solution defines a common reference frame, so we can neglect the small $\frac{v^2}{c^2}$
factors coming from boosting between the centre of mass frames of the different events.
Thus writing 
\f
x^a_K = h_{L b}^{\ a} z_K^a
\ff
for the spatial positions in this common frame, to zeroth order in $g$, the variety is now written in terms of distinctions
used in the non-relativistic real ensemble theory\cite{real2}
\f
{\cal D}_{IJ} \rightarrow {\cal D}_{IJ}^0= \frac{1}{N}
\sum_{K}  \left [ \frac{ x_I^a -x_K^a }{|  x_I^a -x_K^a   |}
-\frac{ x_J^a -x_K^a }{|  x_J^a -x_K^a   |} \right ]^2
\label{D0}
\ff

\subsection{Ensemble approximation}

Let us now consider a large and complex history within which there is a subset of events,
$ {\cal S}$ whose views are mutually similar,
\f
{h}(I,J) <  \epsilon \ \ \ \ \ \  \forall I,J \in {\cal S}
\ff
where $\epsilon << 1$ is a pure number.

We assume in particular that all members of $\cal S$ have the same number of past momenta.

Let ${\cal O}$ be an observable on the space of views, $\cal V$.  We then define its average over the ensemble as
\f
< {\cal O} > = \frac{1}{N} \sum_{K \in {\cal S}}  {\cal O}_K
\ff
Let $w^\alpha$ be coordinates on the space of these views (which in the non-relativistic
approximation is given also by $x^a_K$).  
There will then be a population  density function $\rho (w)= \rho(x^a_K) $ such that
\f
< {\cal O} > = \int d^dw \rho (w) {\cal O}(w) =   \sum_K \int d^dx \rho (x^a_K) {\cal O}(x^a_K)
\ff

\subsection{Consequences of the Principle of the Identity of the Indiscernible}

Recall that the PII dictates that there are no two $I$ and $J$ whose views are identical 
$[p_I^K]_a$, and that this is a consequence of the dynamics we have proscribed.
This means there are no two $I$ and $J$ such that,
\f
[p_I^K]_a = [p_J^K]_a.  
\ff
This means we may regard the view $[p_I^K]_a$ as a function of $I$:
\f
[p_I^K]_a = [p^K]_a (I)
\ff
USing the $PII$ each view can be labeled by its content, so that in the ensemble
approximation
\f
 [p^K]_a (I) =  [p^K]_a (w(I) ) =  [p^K]_a (x^a_I ) 
\ff

\subsection{Cycle identities}

Consider the case where there are two causal paths from one event to a later event.  The simplest case, which is an elementary causal diamond will suffice to make the point.
Here we have 
\f
 z_1 < z_2 < z_4 , \ \ \ \  z_1 < z_3 < z_4
\ff
with $z_2$ and $z_3$ acausal.  Assuming all the lapses are equal to $\cal M$, it follows from (\ref{emergence}) that to zeroth order in $g$,
\begin{eqnarray}
z_4^a - z_1^a & = &  ( z_4^a - z_2^a ) + (z^a_2- z_1^a )  = 
 ( z_4^a - z_3^a ) + (z^a_3- z_1^a ) 
 \\
 && = {\cal M} \left [ p_4^{2 \ a} + p_2^{1 \ a}
 \right ] =  {\cal M} \left [ p_4^{3 \ a} + p_3^{1 \ a}
 \right ]
\end{eqnarray}
More generally, any elementary closed cycle will have a corresponding identity, as a consequence of the consistency
\f
\sum_{cycle} p_a^0 = 0
\label{cycle}
\ff
Note that because of this cycle identity, given any closed curve $\gamma \in {\cal V}$
we have, to zero'th order in $g$,
\f
\oint_\gamma p =
\oint ds p_a^0 (\gamma (s) ) \dot{\gamma}(s)^a  =0 
\ff
from which it follows that there is an $S^0 (x,s) $ such that, 
\f
p_a^0 (s) = \partial_a S^0 (x(s),s) , 
\label{HJ}
\ff

However, to first order in $g$ we also find, suppressing indices.
\f
\oint_\gamma p^1 (x(s))= \oint {\cal F} =\oint \frac{\partial {\cal V}}{\partial x^a} =0
\ff
So through first order in $g$ there is an $S=S^0 + g S^1$ such that
\f
p_a = \frac{\partial S}{\partial x^a}
\ff


\subsection{An ensemble approximation to the action}

We use the cyclic identities to write (\ref{HJ}),
\f
p_a (s) = \partial_a S (x(s),s) , 
\ \ \ \  \dot{p}_a (w) =       \frac{1}{\rho (w)}   \partial_a ( \rho (w) \dot{S} (w) )
\ff


We now compute the ensemble approximation to the action.  As shown in \cite{ECS1}, we have
\begin{eqnarray}
S &=& - \sum_{trajectories}  \int ds x^a \dot{p}_a = 
 \sum_{trajectories}  [ \int ds [ \dot{x}^0 {p}_0  -x^i \dot{p}_i  ]
 \nonumber \\
 &=&  \sum_{trajectories}   \int ds [ -m^2 c^4 + \frac{p(s) _i^2 }{2m} +x^i \dot{p}_i  ]
 \nonumber \\
 \end{eqnarray}
 Now we replace the sum over trajectories of the ensemble with the averages
\f
  S=
 \int_{\cal V} dw  \int ds  \rho [w,s]  
 \left  [ -w^i \frac{1}{\rho} \partial_i \rho (w) \dot{S}(w,s)    - m c^2 -  \frac{(\partial _i  S )^2 }{2m}   \right ]  
\ff

 The first term is
 \f
  \int_{\cal V} dw  \int ds  \rho [w,s]  
 \left  [ -\frac{1}{\rho } w^i \partial_i  \rho (w) \dot{S}(w,s)   \right ] =
 n  \int_{\cal V} dw  \int ds 
 \left  [  \rho (w) \dot{S}(w,s)   \right ]
 \ff
  We next absorb the factor of $n$ into $S$ which is compensated for by a redefinition of the mass
 \f
 m \rightarrow m^\prime= n^2 m .
 \ff
 
 We then have
 \f
 S=  \int_{\cal V} dw  \int ds  \rho [w,s]  \left [    \dot{S}(w,s)    -n^2 m c^2 - \frac{(\partial _i  S )^2 }{2m}     \right ]
 \ff

This is the action for an ensemble of particles with probability density $\rho$
following the gradient of the action $S$ according to the Hamilton Jacobi equation.
The latter is given by the variation by $\rho$
\f
\dot{S} = \frac{1}{2m} (\partial_i S)^2 +m c^2
\label{HJ1}
\ff
while the variation by $S$ gives the law of probability conservation.
\f
\dot{\rho} + \frac{1}{m}\partial_i (\rho \partial^i S )
\ff


\section{Recovery of quantum mechanics}

We next add the variety term to get some dynamics.  
We use the result (eq. 30) from section 3.1 of \cite{real2}, that shows that (\ref{D0}), in the ensemble average has a leading term which is
\f
S^{RE} = -\frac{g}{4}  \int dw ds \rho \ ( \frac{1}{\rho} \partial_i  \rho )^2  + \ldots
\ff
where, as discussed in  \cite{real2}, we have neglected higher order terms in an expansion in derivatives.

We find that we have the real and imaginary parts of the Schroedinger equation, with (\ref{HJ1}) modified by the addition of the quantum potential
\f
\dot{S} = \frac{1}{2m} (\partial_i S)^2 +  g \frac{\nabla \sqrt{\rho} }{\sqrt{\rho}}
\label{rSch}
\ff

The wave function 
\f
\psi = \sqrt{\rho} e^{\frac{\imath}{\hbar}S} e^{-\imath m c^2 t},
\ff
then satisfies the Schrodinger equation,
\f
\imath \hbar \frac{d \psi}{dt}= \left [
-\frac{\hbar^2}{2m} \nabla^2 
\right ] \psi
\ff
with
\f
g=\frac{\hbar^2}{2m}
\ff
We note that there is no potential energy term because we have  taken the limit of a relativistic particle picture in which interactions come from events where particles interact.  It would be impossible to insert a spacetime dependent potential energy into the Hamiltonian constraint because initially there is no spacetime.  The spacetime coordinates $z^a$ emerge from Lagrange multipliers.

\section{Conclusions}

In this paper we have proposed a new approach to fundamental physics, based on a simultaneous completion of quantum mechanics and general relativity, and we have investigated a few of its consequences.  From relative locality and energetic causal sets we take the proposal that what is fundamental in nature are flows of energy and momentum amongst events, defining a causal structure.  Meanwhile, the expression of that causal structure in terms of embedding  into a space-time manifold turns out to be an emergent and coarse grained description.  

We also take two ideas from the real ensemble formulation of quantum mechanics.  First,  the idea that quantum states correspond to ensembles of similar systems, and, second, a proposal for dynamics of those ensembles based on maximizing a measure of their diversity  or variety.  The variety is to be defined by comparing {\it views} of subsystems\cite{variey1}.  In \cite{real1,real2} the view of a subsystem is defined, roughly  as what that subsystem knows about its surroundings as a  result of interacting with it.  Here we define instead the {\it view of an event}  as the set of energy and momentum transferred to it from vents in its immediate causal past.

The main result is the existence of a Newtonian limit within which we are able to derive non-relativistic quantum mechanics as the statistical description of ensembles with large numbers of events.

There is a great deal more to be done to develop this proposal.  To mention just one: the ingredients for extending an energetic causal set model to a discrete spacetime history analogous to a spin foam model, along the lines contemplated in \cite{ECS1,wolfgang} are
present in the connections $[U_I^K ]_a^{\ \ b}$ of \ref{pt} on the causal set and 
$[{\cal O}_I^J]_i^j$ (\ref{DIJ}) in the definition of variety.  We need to investigate proposals for their dynamics.  

Finally, there are implicitions of this proposal for foundational issues such as the measurement problem and the question of physical correlates of qualia.  These will require careful consideration and are beyond he scope of this paper.





\section*{ACKNOWLEDGEMENTS}

Thank you very much to Marina Cortes and Roberto Mangabeira Unger, for their collaboration on the larger project of developing the hypothesis that time is fundamental and irreversible, of which this is a part.  I am also grateful to Henrique Gomes for a careful reading of the manuscript. Finally, thank you to Julian Barbour for decades of conversation.

This research was supported in part by Perimeter Institute for Theoretical Physics. Research at Perimeter Institute is supported by the Government of Canada through Industry Canada and by the Province of Ontario through the Ministry of Research and Innovation. This research was also partly supported by grants from NSERC, FQXi and the John Templeton Foundation.


\end{document}